\documentclass[twocolumn]{article}
\usepackage{lmodern}
\usepackage{preprint}
\usepackage{cite}
\usepackage{amsmath, amsthm, amssymb, amsfonts}
\usepackage{graphicx}
\usepackage{textcomp}
\usepackage{multirow}
\usepackage{array}

\usepackage[utf8]{inputenc}	
\usepackage[T1]{fontenc}	
\usepackage{xcolor}		
\usepackage[colorlinks = true,
            linkcolor = purple,
            urlcolor  = blue,
            citecolor = cyan,
            anchorcolor = black]{hyperref}	
\PassOptionsToPackage{hyphens}{url}\usepackage{hyperref}
\usepackage{booktabs} 		
\usepackage{nicefrac}		
\usepackage{microtype}		
\usepackage{float}			

\usepackage{newfloat}
\DeclareFloatingEnvironment[name={Supplementary Figure}]{suppfigure}
\usepackage{sidecap}
\sidecaptionvpos{figure}{c}

\usepackage{titlesec}
\titlespacing\section{0pt}{12pt plus 3pt minus 3pt}{1pt plus 1pt minus 1pt}
\titlespacing\subsection{0pt}{10pt plus 3pt minus 3pt}{1pt plus 1pt minus 1pt}
\titlespacing\subsubsection{0pt}{8pt plus 3pt minus 3pt}{1pt plus 1pt minus 1pt}

\title{A Survey of Post-Quantum Cryptography Support in Cryptographic Libraries}

\author{
 Nadeem Ahmed\\
 Department of Information Systems\\
 Univ. of Maryland Baltimore County\\
 Baltimore, USA\\
 \texttt{nahmed5@umbc.edu}\\
  \And
 Lei Zhang\\
 Department of Information Systems\\
 Univ. of Maryland Baltimore County\\
 Baltimore, USA\\
 \texttt{leizhang@umbc.edu} \\
 \AND
 Aryya Gangopadhyay\\
 Department of Information Systems\\
 Univ. of Maryland Baltimore County\\
 Baltimore, USA\\
 \texttt{gangopad@umbc.edu} \\
}

\begin{document}

\twocolumn[ 
  \begin{@twocolumnfalse} 
  
\maketitle

\begin{abstract}
The rapid advancement of quantum computing poses a significant threat to modern cryptographic systems, necessitating the transition to Post-Quantum Cryptography (PQC). This study evaluates the support for PQC algorithms within nine widely used open-source cryptographic libraries---OpenSSL, wolfSSL, BoringSSL, LibreSSL, Bouncy Castle, libsodium, Crypto++, Botan, and MbedTLS---focusing on their implementation of the NIST-selected PQC finalists: CRYSTALS-Kyber, CRYSTALS-Dilithium, FALCON, and SPHINCS+. Our analysis, based on the latest available documentation, release notes, and industry reports as of early 2025, reveals a varied state of readiness across these libraries. While some libraries have integrated PQC support or have clear implementation roadmaps, others lag behind, creating potential security risks as quantum threats become more imminent. We discuss key challenges, including performance trade-offs, implementation security, and adoption hurdles in real-world cryptographic applications. Our findings highlight the urgent need for continued research, standardization efforts, and coordinated adoption strategies to ensure a secure transition to the quantum-resistant cryptographic landscape. 
\end{abstract}
\keywords{post-quantum cryptography, quantum threats, open-source cryptographic library} 
\vspace{0.35cm}

  \end{@twocolumnfalse} 
] 


\section{Introduction}
Quantum computing promises to solve certain mathematical problems far faster than classical computers, posing a serious threat to current cryptographic protocols. In particular, Shor’s algorithm, a quantum algorithm, can efficiently factor large integers and compute discrete logarithms---the hard problems that underlie RSA and elliptic-curve cryptography (ECC)\cite{shorPolynomialTimeAlgorithmsPrime1997}. This implies that a quantum computer with sufficient power \textbf{has the potential to compromise all commonly used public-key encryption and digital signature schemes}\cite{NISTsPleasantPostquantum2022}. Grover’s algorithm, another quantum algorithm, can quadratically speed up brute force search, effectively halving the security of symmetric ciphers (for example, AES-128 would have the security of a 64-bit key)\cite{groverFastQuantumMechanical1996,NISTsPleasantPostquantum2022}. Although Grover's impact can be countered by doubling key sizes, Shor's algorithm spells an eventual disaster for RSA, ECC, Diffie-Hellman, and related schemes. The urgency is amplified by the ``harvest now, decrypt later'' threat: adversaries may record encrypted data today and decrypt it in the future once quantum capabilities are available\cite{NISTsPleasantPostquantum2022,levineCurrentStateTransport2024}. Sensitive information with a long shelf life (national secrets, health records, industrial IP) is particularly at risk. The impending threat serves as the catalyst for the development of Post-Quantum Cryptography (PQC), which introduces novel cryptographic algorithms designed to withstand quantum attacks\cite{UnderstandingQuantumThreats2025}.

Over the past decade, PQC has moved from theory to practice. In 2016, the US National Institute of Standards and Technology (NIST) launched an open competition to standardize quantum-resistant cryptographic algorithms\cite{NISTAnnouncesFirst2022}. After three rounds of evaluation, NIST announced a set of finalist algorithms in 2022 to form the core of future standards\cite{NISTAnnouncesFirst2022}. These include \textbf{CRYSTALS-Kyber} (for public-key encryption or key establishment) and three digital signature schemes: \textbf{CRYSTALS-Dilithium, FALCON, and SPHINCS+}\cite{NISTAnnouncesFirst2022}. Kyber and Dilithium in particular are lattice-based schemes, noted for relatively small key sizes and strong performance, whereas SPHINCS+ is hash-based, with larger signatures but with the benefit of a totally different security basis. NIST designated Dilithium as the primary signature algorithm and FALCON as an alternative for applications that require smaller signatures, with SPHINCS+ as a conservative backup\cite{NISTAnnouncesFirst2022}. The selection of these algorithms is a major milestone, but adopting them in real-world systems is an equally critical challenge. In practice, software developers and security engineers must integrate PQC into the protocols and products we use every day. This raises the research question that we aim to answer: \textbf{ Are the common cryptographic libraries on which we rely on ready for the quantum era?}

This research study examines the state of post-quantum algorithm support in widely used open-source cryptographic libraries, with the main focus on implementations of the NIST PQC finalists. The structure of this survey introduces us to the problem in Section 1. In Section 2 we summarize the main objectives and scope of our research. In Section 3 we give a brief background on NIST standardization process, the US government readiness initiatives, PQC algorithms, and existing research on PQC libraries' readiness. Section 4 expands on how we selected open-source libraries and our readiness analysis method. Section 5, our main body of research, reports on the readiness of each of the libraries, Section 6 addresses real-world implementation needs, and Section 7 concludes the study with an outlook on future work needed in open-source cryptographic libraries.

\section{Objectives and Scope}
We evaluate nine popular libraries across consistent criteria and determine which widely used open-source cryptographic libraries have implemented support for the NIST-selected post-quantum algorithms (specifically the NIST PQC finalists and standards). Our definition of a cryptographic open-source library includes any specific software implementation of modern cryptographic primitives and algorithms completed by the open-source community. There are more than 15 such open-source libraries integrated into projects that require cryptographic primitives\cite{ComparisonCryptographyLibraries2025}. For this survey, we focus only on nine that include OpenSSL\cite{OpenSSLLibrary}, wolfSSL\cite{WolfSSLEmbeddedSSL}, BoringSSL\cite{BoringsslGitGoogle}, LibreSSL\cite{LibreSSL}, Bouncy Castle\cite{BouncyCastleOpensource}, libsodium\cite{IntroductionLibsodium2024}, Crypto++\cite{Cryptocpp}, Botan\cite{BotanBotana}, and MbedTLS\cite{MbedTLS} given their wide adoption in software and systems. The focus is confined to \textbf{NIST PQC finalist algorithms}, specifically those chosen for standardization in 2022: \textbf{CRYSTALS-Kyber} (a KEM based on lattices), \textbf{CRYSTALS-Dilithium} (a lattice-based digital signature), \textbf{FALCON} (signature using lattice techniques), and \textbf{SPHINCS+} (a signature leveraging a stateless hash-based method). These represent the primary new standards for public-key encryption and signatures\cite{NISTAnnouncedWinners2022,NISTAnnouncesFirst2022}. In 2020, NIST published the SP 800-208 standard that covered stateful hash-based signature schemes: the \textbf{Leighton-Micali Signature (LMS)} system and the \textbf{eXtended Merkle Signature Scheme (XMSS)}\cite{cooperRecommendationStatefulHashBased2020}. These are also considered PQC algorithms. If a library supports these algorithms, we mention that. Our focus is only on open-source libraries commonly used in practice; we do not survey specialized research prototypes or closed-source products. The analysis reflects the state of these libraries as of early 2025 and draws on the latest available documentation, release notes, academic and industry reports. By surveying these libraries, we can gauge the readiness of the software ecosystem to migrate to PQC and what gaps remain. We maintain a conceptual high-level focus, with the aim of educating decision-makers and developers of the landscape in 2025. Before diving into the library analysis, we provide background on the quantum threat to modern cryptography and the efforts by NIST and government agencies to facilitate a transition to quantum-safe security.

\section{Background and Related Work}

\subsection{NIST PQC Standardization Process}

\paragraph{Motivation and the NIST Competition}
Recognizing the quantum threat, NIST launched a public competition in 2016 to develop and standardize PQC algorithms\cite{NISTAnnouncesFirst2022}. This process mirrored the earlier AES and SHA competitions in openness and rigor. Cryptographers worldwide submitted candidates in various categories (encryption/KEM and digital signatures). The competition had multiple elimination rounds: Round 1 (69 submissions) in 2017, Round 2 narrowed the field in 2019, and Round 3 (2020–2022) featured the finalists. Evaluation criteria included security (resistance to cryptanalysis by both classical and quantum means), performance on diverse platforms (from servers to IoT devices), key and ciphertext sizes, and other factors such as patent status. In July 2022, NIST revealed four algorithms selected for standardization, nominating them as the "winners" of the competition\cite{NISTAnnouncesFirst2022}. For general encryption (e.g., TLS key exchange), NIST selected CRYSTALS-Kyber, noting its small encryption keys and high speed. For digital signatures, NIST chose CRYSTALS-Dilithium, FALCON, and SPHINCS+ \cite{NISTAnnouncesFirst2022}. Dilithium (lattice-based) is recommended as the primary signature scheme due to its efficiency, with FALCON (also lattice-based) offered for applications needing smaller signatures, and SPHINCS+ (hash-based) providing a fallback based on an entirely different hardness assumption. Notably, three of these four use structured lattices as the source of difficulty, while SPHINCS+ uses hash functions. NIST’s announcement also identified four additional algorithms for further analysis in a “Round 4”. These included candidates such as Classic McEliece (code-based encryption) and others that, while not selected in the first batch, could be standardized later for broader security\cite{NISTAnnouncesFirst2022}. 

\begin{figure}[t]
    \centering
    \includegraphics[width=\linewidth]{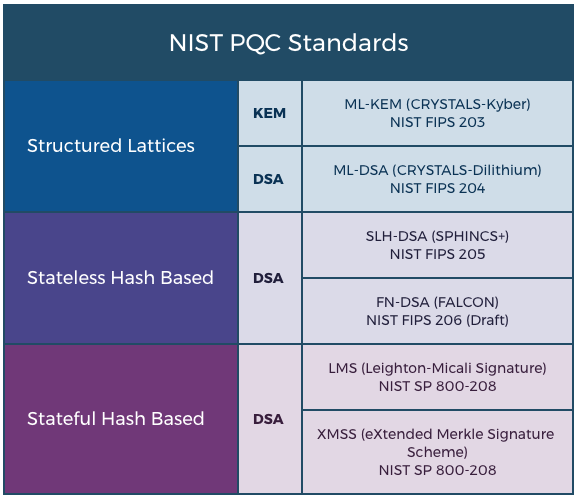}
    \caption{NIST releases final set of encryption algorithm standards\cite{NISTReleasesFirst2024}.}
    \label{fig:NIST Standards}
\end{figure}

\paragraph{Timeline of Standardization}
Following the selection, NIST began drafting standards for the new algorithms. By August 2024, NIST released draft versions of FIPS 203, 204, and 205. These drafts correspond to the standards for Kyber, now called \textbf{Module-Lattice KEM or ML-KEM}, Dilithium as lattice signatures, \textbf{ML-DSA}, and SPHINCS+ as stateless hash signature, \textbf{SLH-DSA}\cite{PostQuantumAlgorithmsOpenSSL2024}. These drafts underwent public comment and final FIPS approval, staying on track with NIST’s earlier estimate that the standards would be finalized about “two years” after the 2022 announcement\cite{NISTAnnouncesFirst2022}. The NIST 2022 press release projected that the post-quantum standards would be finalized around 2024, and by the end of 2024 the principal standards were essentially ready. As of early 2025, the official FIPS documents for Kyber, Dilithium, and SPHINCS+ are being adopted. NIST is expected to continue working on additional standards, including finalizing FIPS 206 (which will be termed \textbf{FN-DSA}), currently in draft, for FALCON and work toward adding other algorithms (e.g., for Classic McEliece or other encryption schemes) in a subsequent round. Fig. \ref{fig:NIST Standards} shows a snapshot of the algorithms chosen by NIST, their equivalent official NIST PQC standard, and the PQC algorithm categories they fall under.

\subsection{Government Readiness Initiatives}
\begin{figure*}[t]
    \centering
    \includegraphics[width=\linewidth]{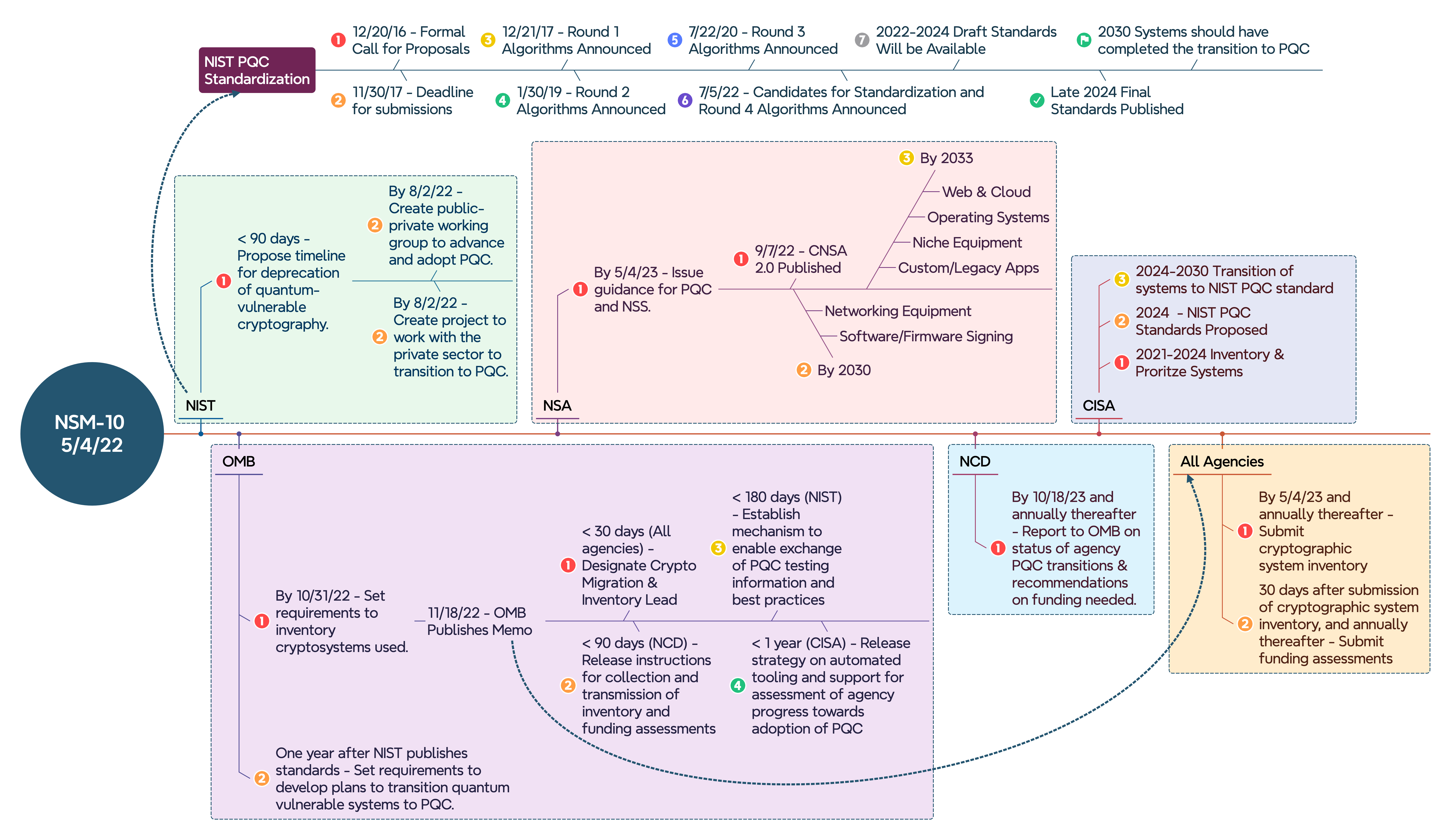}
    \caption{US Government Readiness Timeline showing the origin of government mandates from the release of NSM-10 document which kick starts all other agencies into their own timeline. We also show in parallel where the NIST PQC standardization timeline fits in.}
    \label{fig:Timeline}
\end{figure*}

Other US government agencies have been preparing for this transition in parallel to NIST’s technical work. In May 2022, the White House issued National Security Memorandum 10 (NSM-10), which declared that \textit{“the United States must prioritize the timely and equitable transition of cryptographic systems to quantum-resistant cryptography, with the goal of mitigating as much of the quantum risk as feasible by 2035”}\cite{NSM-10-TechnologySecurity,NationalSecurityMemorandum2022}. NSM-10 tasked agencies to inventory their cryptographic systems and plan for PQC. Subsequently, the US Office of Management and Budget (OMB) released Memorandum M-23-02 (18 Nov. 2022), directing federal agencies (except national security systems) to develop a comprehensive inventory of cryptography in use by 4 May 2023 and update it annually until 2035\cite{OMB-M2302MEMORANDUM}. Agencies must identify where they use public-key algorithms vulnerable to quantum attack (RSA, ECC, DH, etc.) and prioritize systems for transition. The goal is that by 2035 (a timeline based on estimates for practical quantum computers), all vulnerable systems will have been upgraded to quantum-resistant solutions\cite{OMB-M2302MEMORANDUM}. This timeline underscores the 2030s as the deadline for complete migration, with urgent action in the 2020s. Similarly, National Security Agency (NSA) has updated its requirements for National Security Systems (NSS) under its \textbf{CNSA 2.0 suite} (Commercial National Security Algorithm Suite 2.0). In 2022, NSA mandated that certain applications adopt post-quantum algorithms as early as 2025. For example, by 2025, NSS use of software/firmware signing and TLS for web, cloud services must employ quantum-resistant algorithms, and by 2026 the mandate extends to traditional network equipment. NSA’s full guidance calls for all NSS to transition to approved PQC by 2035, in accordance with national policy\cite{PQShield2023,NSAPostQuantum}. Civilian agencies are under similar pressure via the OMB and the Department of Homeland Security (DHS). DHS published a roadmap for the transition to PQC that urges critical infrastructure and industry to start planning inventories and risk assessments now\cite{DHS-PQC,CISA-PQC}. In summary, there is a clear timeline from the top: basic standards by 2024, initial deployments by 2025–2026 in critical areas, and a complete transition by 2035 for US federal systems, with strong encouragement that industry follows a similar schedule\cite{PQShield2023}. This geopolitical push increases the urgency for software libraries and products to incorporate support for PQC at the earliest opportunity, long before the typical developer is compelled to make the transition. Fig. \ref{fig:Timeline} shows us the government initiatives that have their origin in NSM-10 and how they are interrelated in the adoption time frame between US agencies. 

\subsection{PQC Algorithm Categories}
The NIST PQC algorithms span different families of mathematical problems. \textbf{Lattice-based cryptography} yielded the majority of the finalists: Kyber (encryption) and Dilithium/FALCON (signatures) rely on the hardness of lattice problems. These offer efficient operations and relatively small public key sizes (Kyber's public key is around 800-1184 bytes for security levels 1-5, and Dilithium keys are a few kilobytes), which is a big improvement over previous proposals. \textbf{Hash-based signatures} (like SPHINCS+) rely only on cryptographic hash functions; SPHINCS+ produces much larger signatures (tens of kilobytes) but is extremely conservative in security, using no new hardness assumptions. \textbf{Code-based cryptography}, exemplified by Classic McEliece, has very large public keys (several hundred kilobytes) but small ciphertexts and long-standing security. Code-based KEMs were not in the first set of standards, but remain under evaluation. Other categories (multivariate quadratic equations, isogenies) had candidates in the competition (e.g., Rainbow signature, SIKE KEM) but those were ultimately broken or not selected. The selected algorithms thus reflect a balance of lattice and hash based approaches, with an emphasis on performance and confidence in security. For example, Kyber and Dilithium are expected to be the “workhorse” algorithms for most applications due to their efficiency and relatively compact key/signature sizes\cite{NISTAnnouncedWinners2022}. FALCON serves niche scenarios that require smaller signatures (with more complex math involving floating-point sampling), and SPHINCS+ serves as a hedge against any unforeseen breakthroughs against lattice schemes\cite{NISTAnnouncedWinners2022}. Understanding these algorithmic differences is important for our library survey. Each library faces the task of implementing some or all of these algorithms. This involves new data structures (e.g., polynomials for lattice ops, huge keys for McEliece), possibly new API designs (KEMs are a new primitive distinct from classic Diffie-Hellman), and handling performance and memory impacts. Next, we consider existing research on how ready systems are for PQC, and then we delve into each library.

\subsection{Existing Research and Surveys on PQC Readiness}
The transition to post-quantum cryptography has been analyzed in both academic literature and industry reports. Broadly, two threads of research exist: (1) studies of PQC adoption in protocols and systems (e.g., post-quantum TLS experiments, performance measurements, interoperability tests) and (2) surveys or readiness assessments of the crypto ecosystem (standards, libraries, tools) to support PQC. We summarize some key findings from prior work and identify the gaps that our study addresses.

\paragraph{Post-Quantum TLS and Performance}
A number of works have looked at using PQC algorithms in TLS 1.3, the most common secure protocol. A recent comprehensive study (Sosnowski et al., 2023) measured handshake latency and computational costs when swapping PQC algorithms for TLS key exchange and authentication\cite{Sosnowski2023}. They tested finalists like Kyber KEM for key agreement and Dilithium/Falcon for signatures across different libraries and network conditions. A striking result was that post-quantum TLS can be as fast or faster than classical TLS: \textit{“HQC and Kyber are on par with our current state-of-the-art, while Dilithium and Falcon are even faster”} at equivalent security levels\cite{Sosnowski2023}. In other words, replacing ECDH with Kyber, or replacing ECDSA with Dilithium, did not incur a performance penalty in their experiments and at higher security levels (where modern cryptographic algorithms need very large keys or curves), the PQC schemes actually outperformed the traditional ones. They also \textit{observed no significant performance drawback for hybrid modes} (combining modern cryptographic algorithms and PQC), and in some high-security scenarios, “PQC outperformed any [classic] algorithm in use today”\cite{Sosnowski2023}. This challenges a common perception that PQC is slow: At least for the finalists, well-optimized implementations can be efficient. Of course, network latency can dominate handshake cost; the study concluded that PQC is suitable for adoption in today’s systems from a performance perspective\cite{Sosnowski2023}. Another study by Google in real world settings (Chrome experiments) found that hybrid X25519+Kyber key exchanges added minimal overhead, confirming that users do not notice a difference in page load times\cite{GoogleNewPathKyber}. These findings are encouraging: they suggest that libraries that enable PQC will not face insurmountable performance issues, although they must deal with \textit{larger message sizes} (e.g., $\sim$1KB key shares instead of 32 bytes). As an example of size impact, one source notes Kyber768’s public key is $\sim$1088 bytes vs 32 bytes for X25519, and requires more CPU cycles (millions vs tens of thousands for X25519)\cite{levineCurrentStateTransport2024}. But modern networks and hardware can handle this overhead, and techniques such as hybrid key exchange mitigate any single-point weakness. 

\paragraph{Migration Readiness Surveys}
On the broader question of readiness, organizations such as the US Department of Homeland Security and ENISA in Europe have issued guidelines and conducted surveys. DHS’s Post-Quantum Cryptography Roadmap (2021) and the follow-up guidance stress inventory and planning as immediate steps\cite{CISAPQCReadiness2023}. A 2023 joint report by NSA, DHS-CISA, and NIST urged all organizations (not just the government) to begin “quantum-readiness roadmaps” and highlighted the need for crypto agility in products\cite{CISAPQCReadiness2023}. However, these are high-level recommendations. Few published studies systematically examine the software support for PQC. A notable resource is the PQC Capabilities Matrix maintained by the Industry-based PKI Consortium\cite{PQCCapabilitiesMatrix}. This matrix (updated through late 2024) lists software products and libraries and whether they support various PQC algorithms. It provides a quick glance view of support for algorithms like Dilithium, Falcon, Kyber, etc., across many vendors. For example, according to the matrix, Bouncy Castle had checkmarks for essentially all NIST finalist algorithms as of November 2022, whereas OpenSSL (at that time) had none in mainline\cite{PQCCapabilitiesMatrix}. Such resources are helpful, but often lack detail on how the support is implemented and the quality (experimental vs. production-ready). Our research is based on information such as the PQC matrix, diving deeper into the state of each library.

\paragraph{Security Analysis of Implementations}
A critical aspect of readiness is not just having an algorithm implemented, but having it implemented correctly and securely. Post-quantum algorithms are complex, and early implementations have already hit some snags. For instance, in late 2023 researchers discovered “KyberSlash” timing attacks affecting multiple implementations of Kyber KEM. The attack exploited non-constant-time arithmetic in decapsulation, allowing an attacker with precise timing measurements to recover a Kyber private key if they could trigger many decryptions on the same key\cite{KyberSlashAttacksPut,bernsteinKyberSlashExploitingSecretdependent2025}. The KyberSlash vulnerabilities were reported and patched in the official reference code by December 2023, but this incident shows that PQC implementations need rigorous side-channel resistance and auditing. It’s not enough to rely on the algorithm’s hardness; the code must be written with the same care as modern crypto (avoiding leaks via timing, memory access patterns, etc.). Some libraries have a head-start here by leveraging well-vetted implementations (for example, liboqs from the Open Quantum Safe project) or formally verifying parts of the code\cite{OQSOurProject}. However, in general, this is an area where more work is needed. As of 2025, there have not been widely publicized CVEs in mainstream libraries specifically for PQC (aside from issues like KyberSlash which were caught in research). This is in part because many PQC integrations are still new or experimental. One can expect that as adoption increases, so will scrutiny; hence, a goal for the community is to proactively audit and formally verify these algorithms. There are efforts like the Open Quantum Safe (OQS) project, which not only provides implementations but also tries to get expert review on them\cite{PQCCapabilitiesMatrix,OQSOurProject}. Our survey will mention whether a library uses such external projects or their own code. 

\paragraph{Gaps in Research}
While there have been several case studies and experimental rollouts of post-quantum algorithms, a gap remains in understanding the readiness of mainstream cryptographic libraries. Prior work shows that technically, PQC can be deployed today with acceptable performance and in some cases effortlessly to end users (hybrid TLS is already being implemented in browsers). The biggest gaps identified are awareness and integration in products, which is exactly where open-source libraries play a crucial role. Much of the focus has been on protocol-level experimentation (e.g., PQC in TLS) or on specific algorithm performance. But many organizations will rely on libraries like OpenSSL, Bouncy Castle, BoringSSL, or MbedTLS to “bake in” PQC, so that applications can use it via familiar APIs. If a library lags in support, that entire ecosystem (e.g., languages or systems built on it) lags. There has not been a comprehensive survey of how far commonly-used libraries (that developers rely on for cryptography) have come in implementing PQC. Our research addresses this by providing a detailed state-of-the-union for PQC in major crypto libraries, highlighting who is ahead, who is behind, and any challenges encountered in those implementations. We build on prior performance evaluations and integration experiments by examining the current official capabilities of each library. In doing so, we highlight practical issues, such as adoption hurdles that are not the focus of algorithmic research papers, but are crucial for real-world deployment. Our work complements existing surveys by providing an up-to-date snapshot (as of early 2025) of library support, which is valuable for practitioners planning migrations and how that progress can be translated into the crypto libraries that form the backbone of secure applications.

\section{Method}
\subsection{Selection and Ranking of Cryptographic Libraries}
 
Historically, there has been no formal methodology for selecting popular cryptographic libraries and ranking them. The only known work in this area used RSA key patterns in open-source projects to identify the most used library\cite{cryptolib}. Given that our research focuses primarily on the support of PQC within cryptographic libraries, we adhered to a straightforward selection process as follows.
\begin{enumerate}
    \item We examined the top ten programming languages\cite{TopProgrammingLanguages} on GitHub and selected three low-level languages: C, C++, and Java. These languages were chosen because of their typical high performance in cryptographic operations.
    \item Higher-level interpreted languages such as JavaScript or Python were intentionally excluded, as their cryptographic libraries are generally bindings to libraries developed in C or C++.
    \item We referenced the general purpose comparison of cryptographic libraries on Wikipedia and chose nine libraries written primarily in C, C++, and Java for ranking\cite{ComparisonCryptographyLibraries2025}.
    \item Our ranking methodology involved multiplying the number of forks of each library on GitHub by the number of stars attributed to those libraries.
    \item It is noteworthy that some libraries, like LibreSSL or BoringSSL, are mirrors on GitHub because their primary repositories are maintained elsewhere (e.g., BoringSSL is maintained by Google in their own repository and LibreSSL is maintained by the OpenBSD project in their repository).
\end{enumerate}

Our PQC survey builds on the ranking derived from this method and explores the details of each library’s PQC support status.

\subsection{Analysis Approach}

Before proceeding to the library-by-library analysis, we summarize the consistent evaluation criteria we apply to each library as follows.

\begin{enumerate}
    \item Does the library support any of the NIST PQC finalist algorithms (Kyber, Dilithium, FALCON, SPHINCS+)?
    \item If so, since what version are they considered stable or experimental?
    \item If not, are there stated plans/roadmap for adding support?
    \item Do they support alternate PQC schemes (e.g., Classic McEliece, stateful hashes, etc.) outside the main four?
    \item We will clearly note whether support is currently available (in production releases or at least in a development branch) or not.
\end{enumerate}

With these criteria in mind, we now present a detailed survey of the readiness of each cryptographic library for post-quantum cryptography.

\section{PQC Support in Major Cryptographic Libraries}
\begin{figure}[t]
    \centering
    \includegraphics[width=\linewidth]{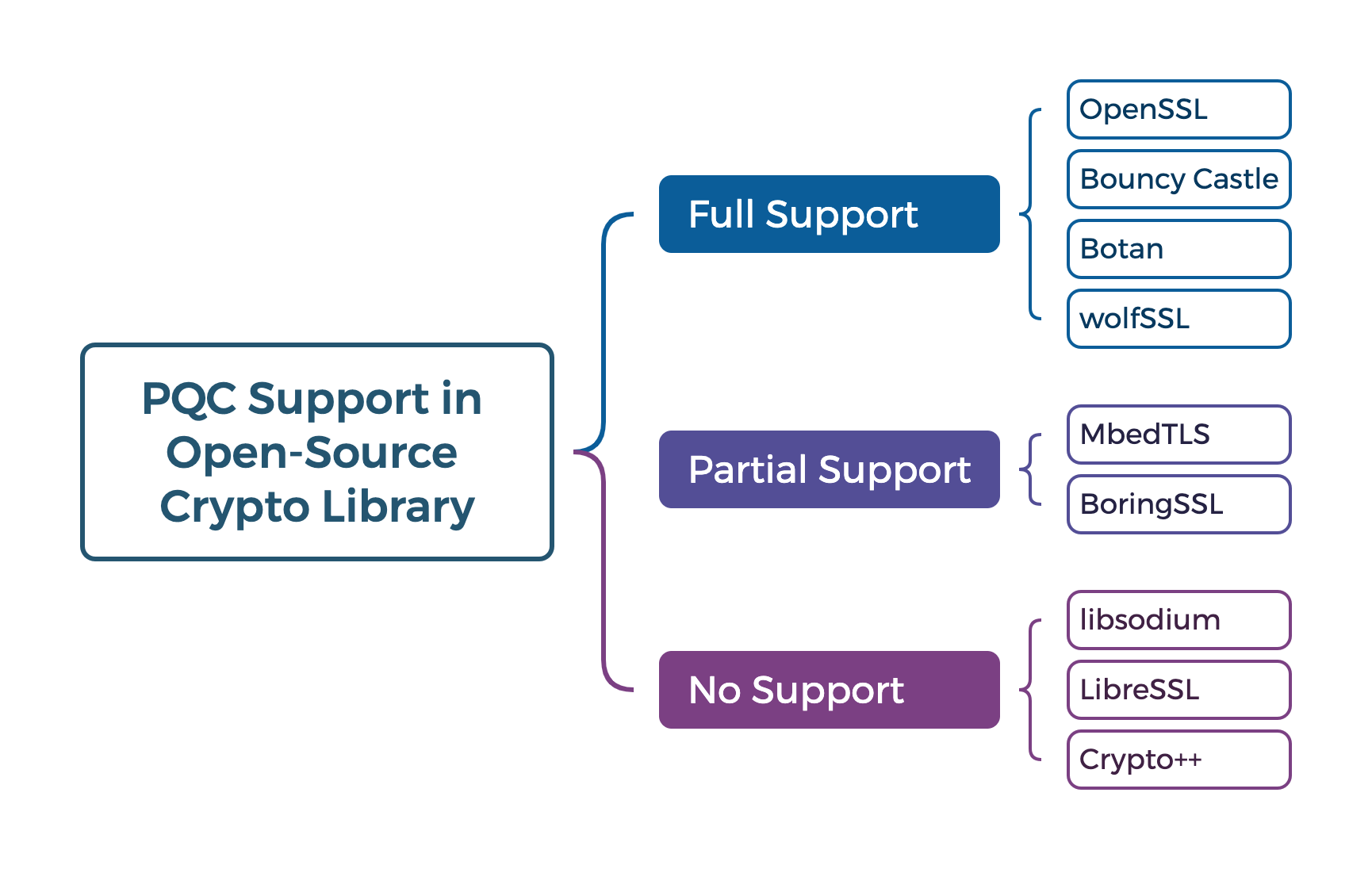}
    \caption{Open-Source Crypto Library PQC Support Summary}
    \label{fig:PQC Support}
\end{figure}

In this section, we evaluate each of the nine target libraries---OpenSSL, libsodium, MbedTLS, Crypto++, Bouncy Castle, wolfSSL, Botan, BoringSSL, and LibreSSL---in terms of their support for NIST’s post-quantum algorithms based on criteria mentioned in previous section. We discuss the current status of implementations and any known challenges or limitations for each library. Toward the conclusion of this section, we encapsulate the support status of the nine libraries in Table \ref{tab:pqc-library-support} and provide an overview of the status of the PQC support in Fig. \ref{fig:PQC Support}.

\subsection{OpenSSL}
As the most widely used open-source cryptographic library, OpenSSL’s stance on PQC is closely monitored. Our ranking places it highest among libraries. As of version 3.4 (2024), OpenSSL has \textbf{not} included NIST PQC algorithms in its official release, but has signaled that support is forthcoming now that standards are finalized\cite{PostQuantumAlgorithmsOpenSSL2024}. In September 2024, the OpenSSL Foundation announced plans to implement the NIST PQC finalists (Kyber, Dilithium, Falcon, SPHINCS+) in a future release, stating: \textit{“We intend to implement support for these algorithms in our providers in a future version of OpenSSL”}\cite{PostQuantumAlgorithmsOpenSSL2024}. This came after NIST published the FIPS draft standards (ML-KEM, ML-DSA, SLH-DSA), which OpenSSL was awaiting to avoid integrating pre-standard algorithms. As of early 2025, integration is underway (targeting OpenSSL 3.5), but not yet released. Meanwhile, since early 2022, the Open Quantum Safe (OQS) project has offered PQC support for OpenSSL 3 via its provider architecture\cite{OQSOurProject}. Maintained by openquantumsafe.org, the OQS provider allowed experimentation with many PQC algorithms across TLS 1.3 and X.509\cite{PostQuantumAlgorithmsOpenSSL2024}, serving as a “test vehicle” for researchers and early adopters. OQS previously offered forks and patches, but the current focus is on the OQS provider. The OpenSSL team supports this initiative; in particular, Michael Baentsch, the principal author of the OQS provider, joined the OpenSSL dev team to assist with native PQC integration\cite{PostQuantumAlgorithmsOpenSSL2024,OQSOurProject}. This collaboration helps OpenSSL adopt PQC with lessons from real-world testing. No alternative PQC algorithms are included in OpenSSL itself yet. Support for RSA/ECDSA with XMSS and LMS exists via engines or patches, not natively. However, OpenSSL's coordination with OQS and public roadmaps demonstrate a strong commitment to PQC readiness, which is vital for infrastructure such as web / email servers and VPNs. Within a year, built-in PQC is expected, enabling seamless use in services like Apache or NGINX using Kyber for TLS 1.3, or verifying Dilithium certificates.

\emph{\textbf{Observation: The OpenSSL PQC support status is currently ``in development, not yet in mainline,'' but through the OQS plugin, PQC can be used with OpenSSL 3.x for testing purposes.}\footnote{In April 2025, OpenSSL officially launched version 3.5, which includes full incorporation of the NIST PQC standard algorithms.}}

\subsection{Libsodium}
Libsodium is a popular cryptographic library known for its simplicity and a consistent API that abstracts low-level details. We rank it second in our list of libraries. It is widely used by developers for ease-of-use in applications ranging from mobile apps to server-side software. As of the latest release in the 1.0.x series, Libsodium does \textbf{not} include support for any of the NIST PQC finalist algorithms. The library currently provides modern public-key cryptography (Curve25519-based key exchange, Ed25519 signatures, X25519/XChaCha, etc.), but no post-quantum key exchange or signature schemes have been integrated yet. The community has shown interest and asked questions regarding PQC in libsodium. Specifically, developers have inquired about the potential addition of algorithms such as Kyber or Dilithium. The lead developer of libsodium (Frank Denis) has been cautious on this front. In a discussion thread, it was noted that “ML-KEM is on the roadmap” for libsodium, but that PQC integration is complex and has not yet been completed\cite{libsodiumPostQuantumCryptography}. This implies that libsodium’s author is aware of Kyber (ML-KEM) and intends to support it eventually, but no official timeline is given. 

\textbf{\emph{Observation: For now, if quantum readiness is required, libsodium alone is insufficient – one must integrate additional tools. This gap highlights that not all libraries have moved at the same pace; ease-of-use libraries such as libsodium may take a bit longer to incorporate cutting-edge PQC algorithms.}}

\subsection{MbedTLS}
MbedTLS (now part of Arm’s Open-Source Security), ranked third on our list and widely used in embedded/IoT, currently has \textbf{no} NIST PQC algorithms implemented in its official release, but it is on their roadmap. The maintainers explicitly stated that they plan to add post-quantum cryptography now that standards are available. A forum query in March 2024 asked about Dilithium/Kyber support, and the response from an Mbed TLS developer was: \textit{"Yes, eventually, we will add standard post-quantum algorithms. We don't know when yet, and we are not going to anticipate NIST standardization. Kyber (as ML-KEM) will likely come first since KEM is more urgent than signature"}\cite{MbedtlsPQCAlgorithms,mbedTLSSupportQuantumSafe}. This confirms that Kyber (ML-KEM) will be implemented first, prioritizing the TLS key exchange need. The developer also noted that for long-term signatures like firmware signing, MbedTLS “already supports LMS"\cite{MbedtlsPQCAlgorithms}. Indeed, MbedTLS added support for stateful hash-based signatures (LMS) around version $\sim$3.2, as part of a contribution (probably motivated by Arm’s interest in secure firmware updates). LMS is a conservative PQC scheme for signatures (stateful, standardized by NIST SP-208). So, while MbedTLS does not yet have Dilithium or Falcon, it does have LMS. This provides one element of quantum-resistance in MbedTLS: an IoT device can sign firmware with LMS, ensuring that signature cannot be forged by a future quantum adversary. However, LMS is not suitable for high-frequency use (stateful, one-time use per signature), and not used in TLS, so it does not solve the handshake problem or general public-key encryption. Thus, MbedTLS still needs to implement the NIST finalists to be broadly quantum-safe. They have not given a specific version or date for when Kyber will land, but the email conversation suggests perhaps 2025 releases might start including\cite{mbedTLSSupportQuantumSafe}. The project might incorporate external code or new code for Kyber. Interestingly, a reply to an email in January 2025 pointed out an implementation called “mlkem-native” (a C90 implementation of Kyber by the PQC Alliance) as a good fit for MbedTLS due to its simplicity and memory safety verification\cite{mbedTLSSupportQuantumSafe}. This suggests that there are already sources of Kyber code that MbedTLS could adopt with confidence. That might accelerate their development.

\textbf{\emph{Observation: MbedTLS does not yet have Kyber, Dilithium, Falcon, or SPHINCS+ implemented, but supports LMS for signatures, and has strong intent to add Kyber soon, followed by Dilithium eventually. This will be a major development because MbedTLS is used in millions of IoT devices (everything from small appliances to sensors).}}

\subsection{Crypto++}

Crypto++ is a widely used C++ library known for its broad set of crypto-primitives. We rank it fourth in our list. Historically, Crypto++ has been cautious in adopting cutting-edge algorithms until standards stabilize. As of version 8.9 (late 2023), it does \textit{not} include the NIST PQC finalists in its mainline distribution. However, third-party efforts have emerged: from 2020–2022, Finnish researchers (J. Hekkala et al.) created a fork adding CRYSTALS-Kyber, CRYSTALS-Dilithium, SABER, and later FrodoKEM\cite{JuliushekkalaCryptopppqc2023,JuliusHekkala2022}. This fork (cryptopp-pqc) proved that PQC integration is feasible and included working code with test vectors, but was never merged into the official project. Crypto++ maintainers likely waited for NIST’s final standards. Although Kyber and Dilithium are now finalized, no official inclusion has occurred as of early 2025. The PQC fork was archived in January 2023 and transferred to the VTT repository (Technical Research Center of Finland), indicating continued work outside the mainline\cite{JuliushekkalaCryptopppqc2023}. Officially, Crypto++ has no PQC support or alternative algorithms. There was a historical McEliece implementation (in 5.x) that was later removed due to lack of maintenance. Crypto++ may be delayed due to its FIPS-focused usage, waiting for certified implementations, or limited resources from a small development team. However, now that NIST has finalized the standards, pressure may increase for native support.

\textbf{\emph{Observation: Crypto++ is not PQC ready in its official release as of 2025. We highlight this as a gap in ecosystem readiness: a popular C++ library missing native PQC support. The hope is that this will change soon given the pressure from standards and users.}}

\subsection{Bouncy Castle}
Bouncy Castle (BC) is a comprehensive crypto library for Java and C\#. Ranked fifth on our list, BC has been notably proactive in implementing PQC. By late 2022, developers stated that “most relevant post-quantum algorithms are now supported” in the API\cite{BouncyCPQCUpdate}. This includes the NIST finalists: Kyber (KEM), Dilithium, Falcon, and SPHINCS+ (signatures), as well as several alternate Round-3 candidates: Classic McEliece, BIKE, HQC, SABER, NTRU, NTRU Prime, and Picnic\cite{BouncyCPQCUpdate}. BC essentially became a playground for PQC algorithms, consistent with its philosophy of early exposure for developers. By version 1.72 (end of 2022), BC had aligned Kyber and Dilithium with third-round specs and updated SPHINCS+. It even included SIKE (later deprecated due to its cryptanalytic break)\cite{BouncyCPQCUpdate}. Throughout 2023-2024, BC refined the PQC support in response to NIST drafts, for example, by adjusting encodings and removing variants such as “Dilithium-AES” and “Kyber-90s”\cite{BouncyCastleCrypto}. On the protocol front, BC’s JSSE added initial PQC key exchange support, such as “ML-KEM in TLS” and a hybrid X25519 + Kyber768 suite (XWing KEM)\cite{BouncyCastleCrypto}, aimed at JDK 21+ which introduced KEM APIs. Thus, BC supports both raw algorithms and PQC use in protocols like TLS and CMS. As of late 2024, BC remains largely in sync with finalized standards, although users should track the latest versions, as changes (e.g. OIDs) have been made during standard finalization\cite{BouncyCastleCrypto}.

\textbf{\emph{Observation: It is safe to say that BC supports all NIST finalists (Kyber, Dilithium, Falcon, SPHINCS+ all present and updated) and a variety of alternatives, making it a one-stop library for PQC experimentation in Java.}}

\subsection{wolfSSL}
Ranked sixth on our list of libraries, wolfSSL (and its underlying wolfCrypt library) has been a leader in early adoption of PQC, especially targeting the embedded systems domain. As of 2025, wolfSSL fully supports certain NIST PQC algorithms in its official release, positioning itself as “the world's first cryptography provider supporting CNSA 2.0 compliance” with PQC\cite{WolfCryptPostQuantum}. Specifically, wolfSSL has implemented Kyber and Dilithium and made them available for use in TLS 1.3, as well as in its general cryptographic API. In wolfSSL terminology, Kyber is referred to as “ML-KEM” (Module Lattice KEM) and Dilithium as “ML-DSA” (Module Lattice Digital Signature Algorithm), matching the NIST nomenclature\cite{WolfSSLUnveilsPost2025}. This support is not just on paper: wolfSSL’s PQC was demonstrated in production contexts. For example, in early 2023 wolfSSL participated in hybrid TLS interoperability tests, and by March 2025 wolfSSL’s press release highlighted their PQC capabilities at the Embedded World conference\cite{WolfSSLUnveilsPost2025}. The algorithms are integrated such that the embedded devices can use PQC \textit{alongside} traditional algorithms. For example, a TLS 1.3 handshake doing an ECDHE-Kyber hybrid, and certificates signed with Dilithium or RSA as needed\cite{WolfSSLUnveilsPost2025}. wolfSSL also supports alternate PQC schemes beyond the NIST finalists: notably, it has implemented stateful hash-based signatures LMS and XMSS (these are used for long-term signature needs, such as firmware signing). In fact, wolfSSL documentation mentions that LMS/XMSS were recently added to address PQC for code signing use cases\cite{ExperimentingPostQuantumCryptography}. These are not NIST's “competition” algorithms (they were standardized separately), but show the comprehensive approach of wolfSSL. wolfSSL announcements do not discuss SPHINCS+, likely due to its slower performance as a stateless hash signature scheme. Currently, wolfSSL has chosen to incorporate LMS/XMSS, which are stateful options. No code-based KEM (McEliece) is supported either, presumably due to key size constraints in embedded. wolfSSL says that it is “post-quantum ready” for both encryption and signatures with the above algorithms. The focus on embedded efficiency makes it stand out: wolfSSL has effectively demonstrated that PQC is not just for big servers; it can run on microcontrollers and even bare-metal environments\cite{WolfSSLUnveilsPost2025}. This aligns with real-world use cases such as IoT devices that will be deployed in the 2020s, but need to remain secure in the 2030s and beyond. 
 
\textbf{\emph{Observation: wolfSSL’s support status is ahead of the curve: Kyber and Dilithium available, plus LMS/XMSS. wolfSSL markets this as fully PQC ready for NSA’s CNSA 2.0 suite\cite{WolfSSLUnveilsPost2025}.} }

\subsection{Botan}
Ranked seventh on our list, Botan is a C++ crypto library known for its flexibility and support for new algorithms. In the context of PQC, Botan has made significant progress and supports the NIST PQC finalists in recent versions. By late 2023, Botan had added implementations of CRYSTALS-Kyber and CRYSTALS-Dilithium to its library\cite{ReleaseNotesBotan}. It also implemented SPHINCS+ in an earlier release \cite{ReleaseNotesBotan}. According to the PQC matrix, Botan supports Kyber, Dilithium, SPHINCS+, and has XMSS already (that was added around Botan 2.x\cite{PQCCapabilitiesMatrix}. Falcon was \textit{not} yet supported at the time of the update\cite{PQCCapabilitiesMatrix}. However, Botan has it on the roadmap for later use. Botan is also considering incorporating support for Classic McEliece. The PQC matrix marks McEliece as planned, implying that the project maintainers are contemplating its inclusion, possibly pending a decision from NIST\cite{PQCCapabilitiesMatrix}. Additionally, Botan developers were active in standardization; the lead maintainer Jack Lloyd often tracks NIST drafts. For example, in Botan’s release notes, they mention deprecating Kyber-90s variant when NIST dropped, and the renaming of enums to mark Round-3 vs. final version\cite{ReleaseNotesBotan}. This shows that they keep their PQC in sync with standards. So effectively, Botan 3.x supports PQC algorithms: Kyber (all security levels), Dilithium (multiple levels), SPHINCS+ (probably one or more set of parameters), plus stateful XMSS. No mention of LMS, but XMSS is there (Botan was one of the first to add XMSS after RFC8391). It is likely that when NIST finalizes Falcon, Botan might incorporate it too, completing the finalist set.  Botan’s support status is quite good, not as comprehensive as Bouncy Castle but covering key standardized ones. Hence, Botan is fairly PQC-ready, making it one of the more prepared C++ libraries. Compared to Crypto++, Botan has clearly moved faster on PQC. Compared to OpenSSL, it is about on par in terms of what is implemented (in fact, ahead, since OpenSSL has not published support yet). The difference is that Botan is less ubiquitous than OpenSSL, but within its user base (which includes some security-focused apps and possibly some government or open-source projects), this support is crucial. 

\textbf{\emph{Observation: With Kyber, Dilithium, and SPHINCS+ implemented and planned support for others, Botan positions itself as a strong option for developers who want to incorporate PQC in C++ applications.}}

\subsection{BoringSSL}
BoringSSL is Google’s fork of OpenSSL, used in Chrome/Chromium and across Google infrastructure. Although not prominent on GitHub, it ranks eighth on our list because of its internal use at Google. Google has long been a leader in post-quantum experimentation through CECPQ experiments\cite{CECPQ22025}, using BoringSSL as the testing platform. In September 2024, Google’s Chrome Security team confirmed that ML-KEM was implemented in BoringSSL\cite{GoogleNewPathKyber}, following Kyber's standardization. BoringSSL previously experimented with KEMs such as NTRU-HRSS and SIKE in CECPQ2, although SIKE was later dropped\cite{CECPQ22025}. Chrome had previously tested a hybrid X25519+Kyber exchange using an experimental IETF codepoint (0x6399)\cite{GoogleNewPathKyber}, later replaced by the official one (0x11EC) after the finalization of ML-KEM. As of Chrome version 131, standardized ML-KEM fully replaced earlier versions. Google began rolling this out to all Chrome Desktop clients\cite{GoogleNewPathKyber}, marking a major real-world PQC deployment. On the server side, Google controls both ends, enabling hybrid support at scale. Cloudflare also adopted BoringSSL and deployed X25519+Kyber to origin servers\cite{CloudflareNowUses2023}, showing broad adoption. For signatures, BoringSSL includes code for PQC signature verification if needed, and there is test support for hybrid ECDSA + Dilithium\cite{CiC-1-2-6}. As of early 2025, BoringSSL also implemented Dilithium (ML-DSA) directly for signatures. Although not widely used outside of the Google and Chromium projects, the success of BoringSSL has influenced other libraries and helped to demonstrate that PQC can scale without disruption. Its contributions have also been fed into OQS efforts via BoringSSL forks\cite{CiC-1-2-6}.

\textbf{\emph{Observation: BoringSSL is PQC ready, at least partially, and in production just for Kyber and Dilithium, making Google one of the first to protect real user traffic with post-quantum algorithms.}}

\subsection{LibreSSL}
Ranking ninth on our list is another internal library that powers much of the open-source infrastructure, especially operating systems derived from OpenBSD. LibreSSL is an OpenSSL-derived library maintained by the OpenBSD project that focuses on simplicity and security. As of 2025, LibreSSL does \textit{not} have support for NIST post-quantum algorithms in any released version. The project has not announced any roadmap or intention with respect to PQC, at least not publicly in mailing lists or release notes. In the LibreSSL 3.7, 3.8, 3.9 release notes (2022–2024), there is no mention of Kyber, Dilithium, or related OIDs. This is perhaps unsurprising. LibreSSL historically lags behind OpenSSL in adding new features (for example, TLS 1.3 and many modern ciphers took a while to appear in LibreSSL). The LibreSSL team tends to prioritize cleaning up code and maintaining security of existing features rather than adding experimental ones. Given that OpenSSL itself had not yet integrated PQC, LibreSSL had no upstream implementation to pull in. We can infer that they are \textit{waiting for the dust to settle}. OpenBSD and most other flavors of BSD Unix including Darwin (from which Apple's macOS is also derived) use LibreSSL. OpenBSD has experimented with hybrid ECDH+Kyber in their current branch for SSH in late 2023 (there were reports that OpenSSH 9.3 added support for X25519+Kyber for experimentation). That is in OpenSSH (which uses its own crypto or calls OpenSSL), not in LibreSSL per se. If OpenSSH has a bundled PQC algorithm, it might rely on built-in code or the OQS library. The PKI consortium matrix did not list LibreSSL at all, presumably indicating that there are no PQC features to report\cite{PQCCapabilitiesMatrix,LibreSSL384391}. It is possible that OpenBSD will integrate some PQC via their signify tool or so for signing, but again, not clearly through LibreSSL. 

\textbf{\emph{Observation: Users of LibreSSL do not have an out-of-the-box quantum-resistant option via that library. This lag could become a concern as the government and industry push for quantum readiness. So, as it stands: LibreSSL’s support status is “none yet.”}}

\begin{table}[t]
\centering
\caption{Open-Source Crypto Library PQC Support Matrix}
\label{tab:pqc-library-support}
\begin{tabular}{>{\hspace{0pt}}m{0.145\linewidth}>{\centering\hspace{0pt}}m{0.103\linewidth}>{\centering\hspace{0pt}}m{0.196\linewidth}>{\centering\hspace{0pt}}m{0.125\linewidth}>{\centering\arraybackslash\hspace{0pt}}m{0.175\linewidth}} 
\toprule
\textbf{Library} & \textbf{Language} & \textbf{Supports NIST PQC} & \textbf{PQC Forks Exist} & \textbf{NIST Standards Roadmap}  \\ 
\midrule
OpenSSL          & C                 & Yes                         & Yes                      & ---                                     \\ 
\midrule
libsodium        & C                 & No                         & No                       & No                                      \\ 
\midrule
MbedTLS          & C                 & Partial                         & Yes                      & Yes                                     \\ 
\midrule
Crypto++         & C++               & No                         & Yes                      & No                                      \\ 
\midrule
Bouncy Castle    & Java / C\#              & Yes                        & No                      & ---                                     \\ 
\midrule
wolfSSL          & C                 & Yes                        & No                      & ---                                     \\ 
\midrule
Botan            & C++               & Yes                        & No                      & ---                                     \\ 
\midrule
BoringSSL        & C                 & Partial                    & Yes                      & Yes                                     \\ 
\midrule
LibreSSL         & C                 & No                         & No                       & No                                      \\
\bottomrule
\end{tabular}
\end{table}

\section{Real-World Use Cases and Implementation}

The drive towards post-quantum cryptography is not occurring in isolation; it is motivated by specific applications and needs across different fields. We highlight a few key areas where PQC and the support in the above libraries are being applied or evaluated.

\paragraph{Government and Military Systems}
Government agencies (civilian and military) often handle data that must remain confidential for decades (diplomatic communications, intelligence, etc.\cite{PQShield2023}. These organizations are among the first to mandate quantum-resistant measures. The US government, for example, through the NSA’s CNSA 2.0 suite, has stated that national security systems should start using approved PQC algorithms by 2025 for certain applications\cite{PQShield2023}. An example is software/firmware signing for military equipment: a fighter jet or satellite launched in 2025 may be in service until 2050, so its software updates must use signatures (like Dilithium or LMS) that a quantum computer in 2040 cannot forge. We see libraries responding: wolfSSL and MbedTLS adding LMS specifically for this use\cite{BlogWolfSSL2024,MbedtlsPQCAlgorithms}. Government secure communications (e.g., VPNs between agencies) are also being tested with PQC. OpenVPN and SSL/TLS/IPsec stacks in government use (often OpenSSL) are expected to adopt hybrid key exchanges as soon as OpenSSL supports them\cite{PostQuantumAlgorithmsOpenSSL2024}. In addition, government-regulated critical infrastructure (energy grids, transportation) is being advised by agencies such as Cybersecurity and Infrastructure Security Agency (CISA) to inventory and upgrade crypto\cite{CISAPQCReadiness2023}. These often use libraries like OpenSSL/MbedTLS in their control systems, so the readiness of those libraries directly impacts these sectors. For instance, a power grid sensor network using MbedTLS will likely move to a firmware signed with LMS and TLS transport with hybrid ECDH+Kyber once available. Governments occasionally develop their own implementations; however, they are increasingly dependent on the open-source community for reasons of transparency and budget, making the efforts in OpenSSL and similar projects crucial. We see coordination in bodies such as NIST's NCCoE (National Cybersecurity Center of Excellence), which hosts projects and drafts guidelines on PQC migration for enterprises\cite{levineCurrentStateTransport2024}. 

\textbf{\emph{Observation: Government-driven use cases are ensuring that libraries implement PQC sooner rather than later, because compliance dates are being set. Libraries such as wolfSSL already claim to be CNSA 2.0 ready for government use\cite{WolfSSLUnveilsPost2025}. The upcoming support for OpenSSL is also largely driven by government and industry compliance needs\cite{PostQuantumAlgorithmsOpenSSL2024}.}}

\paragraph{Web and Internet Security (Browsers and Cloud)}
Perhaps the most visible real-world deployment of PQC to date is in web browsers and cloud services. In late 2022 and into 2023, Google began deploying a hybrid X25519+Kyber key agreement in Chrome (with BoringSSL). By late 2024, Chrome had this enabled for all desktop use, making it likely the largest post-quantum cryptography deployment ever\cite{GoogleNewPathKyber}. This achievement, executed seamlessly without user disruption, highlights careful engineering and demonstrates the feasibility of PQC at a large scale. Likewise, Cloudflare, which handles a significant fraction of Internet traffic, announced that it “now uses post-quantum cryptography to talk to your origin server”\cite{CloudflareNowUses2023}. Cloudflare integrated PQC (using BoringSSL and its own Go-based CIRCL library) in connections between their servers and customer servers. This kind of real-world example shows that open-source libraries (BoringSSL in this case, and CIRCL which is partially based on OpenSSL’s OQS) are up to the task. On the server side, many companies are preparing: Amazon, Microsoft, and others have also tested PQC in TLS and VPNs. For example, AWS KMS (Key Management Service) has an option for hybrid post-quantum TLS in some regions\cite{AWSPostquantumCryptography2024}. Meta (Facebook) wrote about their internal adoption: they have started using hybrid post-quantum TLS in their data centers to secure traffic between data center systems that might be recorded and later attacked. Meta chose Kyber768+X25519 and even considered using Kyber512 in lower-security internal links to save performance. They first emphasized the focus on susceptible channels with SNDL (store now decrypt later)\cite{PostquantumReadinessTLS2024}. All this was built on libraries: likely they used Fizz (Facebook’s TLS library) which integrated liboqs or BoringSSL. The open source community is often involved; Meta's work helps OpenSSL and others know that it is viable. Another Web use case is Content Signing and Code Signing: For instance, browsers might one day require that software updates or extensions be signed with a PQC algorithm. Windows and Linux update packages in the 2030s probably need PQC signatures. Libraries like Bouncy Castle enabling Dilithium signatures means developers can start building that into update frameworks now. The Web PKI (Certificate Authorities) are also preparing: some CAs have issued test Dilithium and Falcon certificates (DigiCert, for example). They rely on libraries like OpenSSL to parse them. The IETF has drafts for X.509 identifiers for PQC (to ensure interoperability), and open-source libraries are implementing those\cite{BouncyCastleCrypto}. 

\textbf{\emph{Observation: We expect in a few years that all HTTPS connections might be to a website with TLS 1.3 using PQC algorithms for both encryption and certificate verification thanks to PQC integrations.}}

\paragraph{Enterprise and Cloud Infrastructure}
Beyond browsers, enterprises are thinking of internal VPNs, secure email (S/MIME), and storage encryption. Companies such as Microsoft have announced that “quantum-resistant cryptography is here” for some of their products. For example, a VPN in Windows might use an ECDH+Kyber algorithm (Microsoft contributed to the IETF draft for that)\cite{MicrosoftsQuantumresistantCryptography}. In the cloud, Kubernetes clusters might use PQC in their service mesh. The libraries we discussed are the foundation for many of these enterprise systems. For instance, a Java enterprise app can use Bouncy Castle to send a post-quantum TLS request to a server. A C++ microservice might use Botan or OpenSSL with OQS to ensure that only quantum-safe cipher suites are negotiated. Cloud providers have started offering PQC as a service: e.g., AWS CloudHSM offers some level of PQC support (the PKI consortium matrix lists some HSMs supporting Dilithium, Kyber in beta)\cite{PQCCapabilitiesMatrix}. These HSM firmware likely use an embedded library (maybe a stripped down wolfSSL or custom code). But for software, open-source libraries are key---they allow integration into existing products easily. 

\textbf{\emph{Observation: Enterprise use cases often emphasize hybrid deployment (to avoid breaking compatibility). So libraries supporting composite certificates (multiple signatures on one cert) or hybrid key exchange help enterprises phase in PQC without forklift upgrades. Bouncy Castle and WolfSSL provide exactly those capabilities\cite{wolfSSLPostQuantum2024,BouncyCPQCUpdate}.}}

\paragraph{IoT and Embedded Systems}
IoT is sometimes cited as both a challenging area for PQC (due to constraints) and a critically needed area (because the IoT devices deployed now will be around for a long time with potentially sensitive roles)\cite{liuPostQuantumCryptographyInternet2024,officeIntegrationPQCTLS2024}. Real-world examples here include: a smart meter network for utilities---they need to ensure that communications (which may control power switching) are secure against future threats, since meters might be in use for 15+ years. They may use MbedTLS or wolfSSL in their communication modules. We see wolfSSL heavily targeting this domain, showcasing PQC working on microcontrollers and even providing a demo of post-quantum firmware signing with Winbond secure flash memory\cite{BlogWolfSSL2024}. In automotive, car electronic control units often use wolfSSL or MbedTLS for diagnostic communication and updates. On the industrial side, PLCs and SCADA equipment (often running embedded Linux with OpenSSL) will need PQC especially if they have remote update or remote command capabilities\cite{SajimonPQC2022}. We already have an example in the OpenSSH world: OpenSSH 9.3 added support for an experimental hybrid ECDH-Kyber key exchange (using liboqs) for those who want quantum-safe SSH now. Many embedded systems use OpenSSH (which uses OpenSSL), so once OpenSSL has PQC, even embedded devices with SSH can easily turn it on. The challenge in IoT is updating: Many devices will not be updated to support PQC in time, leaving them potentially vulnerable to capture-now-decrypt-later of their traffic. Forward-looking organizations are thus specifying that any new IoT deployments should use PQC-capable libraries so that even if not turned on today, they can enable it via configuration updates in a few years. The idea of "hybrid IoT networks" also exists, in which the gateway might handle advanced post-quantum cryptography, while the small sensors utilize basic symmetric cryptography. However, if the gateway is eventually compromised, it could pose a significant issue. Better to have end-to-end PQC if possible. Efforts such as the Linux Foundation's PQCrypto project (mentioned in the context of wolfSSL) aim to produce customized implementations for embedded\cite{mbedTLSSupportQuantumSafe}.

\textbf{\emph{Observation: Having libraries such as MbedTLS and wolfSSL that primarily target IoT and embedded devices implement PQC early is critical for the IoT sector not to be left behind.}}

\paragraph{Recommendations}
In all these examples, one can see a common pattern: the need for standards compliance, interoperability, and performance. Open-source cryptographic libraries are the linchpin connecting academic algorithms to real-world security protocols. Our survey of OpenSSL, libsodium, wolfSSL, BoringSSL, LibreSSL, Bouncy Castle, Crypto++, Botan, and MbedTLS shows a spectrum of readiness. Some (wolfSSL, BoringSSL, Bouncy) are already being used in production for PQC; others (OpenSSL, Botan) are nearly there, actively integrating; and a few (libsodium, LibreSSL, Crypto++) lag behind, which could limit their use in forward-looking projects. The good news is that, as of 2025, any organization or developer who needs to deploy post-quantum cryptography can find a suitable open-source library to do so. For example, if tomorrow a government agency insists that all VPN connections use a post-quantum algorithm, one could compile OpenVPN against OpenSSL+OQS or wolfSSL and be compliant. If an embedded device manufacturer wants to sign firmware with Dilithium, they can use Bouncy Castle (Java) in their backend to generate keys and wolfSSL on device to verify, or soon use MbedTLS on device if LMS is acceptable. The pieces are falling into place.

To accelerate progress, we recommend the following:
\begin{itemize}
  \item Developers and organizations should start experimenting with PQC now using the libraries that support it or the available forks.
  \item Early testing in specific applications (even in hybrid modes) should be performed to highlight issues and drive refinements.
  \item The community should continue to contribute implementations and optimizations to open-source projects~\cite{PostQuantumAlgorithmsOpenSSL2024}.
  \item More comprehensive benchmarking and security analysis of these library implementations in real-world scenarios (covering throughput, latency, and side-channel safety) should be conducted to ensure production readiness.
  \item Cryptographic library maintainers should treat PQC support as a high priority in their roadmaps over the next 1 to 2 years, given that NIST has finalized the first PQC standards.
\end{itemize}
The “day when quantum computers arrive” is uncertain, but the response of the open-source crypto community, as evidenced by this analysis, is well underway. By continuing on this path, and addressing the gaps identified, we can achieve a state where widely used crypto libraries provide drop-in quantum resilience, enabling the secure systems of today to protect against the threats of tomorrow.

\section{Conclusion}

The analysis of these nine cryptographic libraries reveals a landscape of uneven progress toward post-quantum cryptography support. On one end, libraries like wolfSSL, Bouncy Castle and Botan have moved aggressively to implement NIST’s post-quantum algorithms, providing developers with ready-to-use tools for quantum-resistant encryption and signatures. wolfSSL, Bouncy Castle integrate Kyber and Dilithium into their TLS stack today, and Botan offers a rich PQC suite in its C++ API\cite{wolfSSLPostQuantum2024,BSIProjectDevelopment}. On the other end, libsodium, has yet to embrace PQC, reflecting caution and the challenges of adding these complex schemes to a minimalist library\cite{libsodiumPostQuantumCryptography}. Similarly, PQC support for Crypto++ and LibreSSL is either not yet on their roadmap, or for MbedTLS the conversation has just started.  OpenSSL\footnote{In April 2025, OpenSSL officially launched version 3.5, which includes full incorporation of the NIST PQC standard algorithms.}, the most critical library for Internet security, is in a transitional phase: it does not officially support PQC as of early 2025, but clear plans and ongoing work signal that Kyber, Dilithium, and others will be integrated in the near future\cite{PostQuantumAlgorithmsOpenSSL2024}. In the meantime, OpenSSL users must rely on external providers such as liboqs for PQC capabilities. BoringSSL (Google's fork), although not intended for general distribution, has proven in practice that post-quantum algorithms can be deployed at scale. Chrome’s adoption of hybrid X25519+Kyber key exchange stands as a milestone for real-world PQC use\cite{ProtectingChromeTraffic}.

Collective efforts in these libraries underscore an important point: crypto agility. The community is preparing not just to swap in specific algorithms, but to remain flexible in case new information arises (for example, if a lattice algorithm gets broken, the availability of hash-based or code-based alternatives in libraries provides a fallback). This agility is facilitated by the groundwork these libraries have laid. Our comprehensive study indicates that, while there are challenges ahead, the open-source crypto ecosystem is rising to meet the quantum threat. The finalists of the NIST PQC project are steadily being woven into the fabric of the Internet’s security libraries, which means the tools to protect data in the quantum age are becoming readily accessible to developers across all sectors.



\end{document}